\documentstyle[epsfig]{aipproc}

\begin{document}

\title{             
 Two-Proton Emission in the Hyperharmonics Approach} 
\vspace{-2 mm}

\author{ 
Ivan G.~Mukha 
}     
\vspace{-2 mm}

\address{\vspace{-1 mm} 
Institut f\"{u}r Kernphysik, Technische Universit\"{a}t,
   D-64289 Darmstadt, Germany\\
Associate at Gesellschaft f\"{u}r Schwerionenforschung (GSI), 
 D-64291 Darmstadt, Germany\\
On leave from Kurchatov Institute, RU-123182 Moscow, Russia}

\maketitle
\vspace{-10 mm}

\begin{abstract}
Nuclear decays into three-particle channels are considered
 in a few-body approach of hyperspherical harmonics with emphasis on 
  simultaneous, or direct, emission of two protons.
General conditions  of direct decays  are described and
their main features, being experimentally established in decays of 
light nuclei,
are reported.

The analysis method  based on an expansion of decay amplitude
into a series of hyperspherical harmonics  is reviewed.
The basis of hyperspherical harmonics functions
is a generalisation of the spherical function basis 
in  three-body systems.
The method is tested on analysis of the direct decay $^{6}$Be$\rightarrow\alpha$+p+p 
where the  three-body components in the nuclear structure of $^{6}$Be have been studied.
In particular,  the observed strong proton-proton 
correlations are treated as a manifestation of a specific three-body quantum 
effect: the kinematic focusing of fragments over momenta and in space.

The hyperspherical harmonics method is  applied
for the predictions of proton-proton correlations and life-time estimates of
the nuclei $^{19}$Mg, $^{34}$Ca and $^{48}$Ni - candidates for
 two-proton radioactivity.  Each direct 2p-decay should result in a set of
 peaks in the 
 E$_{p-p}$ spectrum whose number and positions depend on the structure of 
 initial nucleus, 
 opposite to  the 
diproton model,  predicting the $^2$He emission with one peak at E$_{p-p}\approx$0
in all cases.

\end{abstract}
\vspace{-5 mm}

\section*{Direct three-particle decays}
\vspace{-2 mm}

 A simultaneous emission of two protons is
a  genuine three-particle nuclear decay and  is a complementary  mode to the known
sequential emission of protons via narrow  intermediate states.
In general, a sequential mechanism of three-particle decay is  a chain of two
independent binary
decays via a narrow intermediate state whose width should be much smaller 
than the decay
energy. The total decay amplitude is then a product of two binary amplitudes.
When narrow intermediate states are absent the sequential decay mechanism
is not plausible because it contradicts  
the uncertainty and  causality principles. Such non-sequential  decays are 
called  direct decays.    

In  sequential decays  
information about correlations of  fragments in the initial nucleus is lost 
because of strong final state interactions. 
In  direct decays the fragment distributions  are 
not  distorted significantly by  final state interactions and therefore 
reflect correlations of the  respective clusters  in 
 the initial nucleus.
In comparison with the two-particle case,  three-particle decays are
 more informative due to  additional degrees of freedom where 
more observables ({\it eg} energy spectra of 
fragments, correlations between fragments) are available.
Thus,  a direct three-particle decay 
is a promising tool to study   nuclear structure.

Main features of  direct decays   are  experimentally
established in the two-proton decays of the $^6$Be 
ground and first excited states and their analogs in $^6$He and 
$^6$Li$^*$(T=1), (see 
\cite{boch89} and references there). Strong 
nucleon-nucleon 
correlations are observed which are not connected 
with the p-p or $\alpha$-p final state  interactions. 
In particular, the measured energy spectra of $\alpha$-particles 
from the mentioned decays 
display sharp peaks over broad pedestals.
These peaks correspond to
 a strong nucleon-nucleon energy  
correlation which was first interpreted as emission of  $^2$He, or di-proton,
while the pedestals were associated with a sequential proton emission via $^5$Li
\cite{gee77}.
However, such a model fails to explain the angular $\alpha$-p correlations from the
$^6$Be decay measured in the kinematical complete experiment \cite{boch92}.
\vspace{-4 mm}

\section*{Hyperspherical harmonics method}
\vspace{-2 mm}

  An adequate approach  describing  few-body nuclear
interactions is suggested by L.M.~Delves, \cite{delves60}, who 
introduced the hyperspherical harmonics, or K-harmonics, basis which gives
 correct angular wavefunctions like the spherical harmonics  
in the two-particle case. 
Usage of  K-harmonics  makes it
possible to write the asymptotic of the total wavefunction along three-particle
channels in a way which is formally identical with that for two-particle channels.
The way to solve the Schr\"{o}dinger equation for three particles, which
decouples the total wavefunction into radial and angular parts is 
 illustrated below and uses the two-particle case for comparison.

The system of three particles {\em i, j, k}  with total
 energy $E$ can be characterized by 5 kinematical variables\footnote
{All particles are assumed to be spinless,
the general case including  spins is considered in \cite{delves60}.
}.
Instead of the radius in the 2-particle centre of mass system, 
two Jacobi radii, {\bf x} and {\bf y},
 are  introduced 
where one radius is between  particles {\em j} and {\em k}, 
{\bf x}={\bf r}$_j$-{\bf r}$_k$,
 and the another  radius is between the third particle {\em i} and c.m. of 
 the selected pair, 
{\bf y}=(m$_j${\bf r}$_j$+m$_k${\bf r}$_k$)/(m$_j$+m$_k$)--{\bf r}$_i$. 
The respective Jacobi momenta {\bf p}$_x$, {\bf p}$_y$, Jacobi orbital momenta 
${\bf \ell}_x$, $\ell_y$, and Jacobi energies $E_x$, $E_y$  
are defined by analogous 
relations (naturally, 
$E_x$+$E_y$=$E$). Then the 5 kinematical variables may be
${\bf \Omega}_{i}= ({\bf \Omega}_{x},{\bf \Omega}_{y},x_{i})$: the directions
${\bf \Omega}_{x}$ and ${\bf \Omega}_{y}$ of the Jacobi momenta  and the quantity 
$x_{i}$=$\arctan(\sqrt{E_x/E_y})$ which reflects the energy distribution between 
the Jacobi subsystems.

By introducing the hyperspherical coordinates, hyperradius 
$\rho^2$=$r_x^2$+$r_y^2$ and hyperangle
$\theta$=$\arctan{(x/y)}$, 
 one may use
 instead of the orbital operator $\hat{L}$,  the grand orbital operator $\hat{K}$,
which has  eigenfunctions
as the functions of  hyperangle: 

\hspace{3. cm} $\Psi^{l_x,l_y}_{K}\sim\sin^{l_x}(\theta)\cos^{l_y}(\theta)
P^{l_x+0.5,l_y+0.5}_{n}(\cos2\theta),\!\!\!$   

\noindent where $P^{\alpha,\beta}_{n}$ are the Jacobi polynomials.

The respective quantum number is called
hypermomentum, which minimal value is equal to the sum of the Jacobi orbital momenta,
K=$\ell_x$+$\ell_y$+n (n=0,1,2,...).
This additional quantum number 
is a three-body analog of the orbital momentum value. 

  With the $\hat{K}$ eigenfunctions,  one may obtain the solution of 
the three-particle Schr\"{o}dinger equation 
(T+V-E)$\Psi^T_{JM}$=0, with the sum of binary potentials V=V$_{ij}$+V$_{jk}$+V$_{ki}$,
 in a form of a hyperradial
part of a total wavefunction coupled with the functions of a hyperangle:

\hspace{3.5 cm} 
$\psi_{JM} =
\sum{R_{KL}^{l_xl_y}(\kappa\rho)Y^{l_xl_y}_{KLM}({\bf \Omega}_i)}$,\\ 
here 
the hyperspherical harmonics (HH) functions, or K-harmonics, are
\[
 Y^{l_xl_y}_{KLM}({\bf \Omega}_i)=\Psi^{l_xl_y}_{K}(\theta)
\left[Y_{l_x}({\bf \Omega}_x) \otimes Y_{l_y}({\bf \Omega}_y) \right] ^{LM},   
\]
[...]$^{JM}$ denotes the {\bf L+S} vector addition to form {\bf J}. The K-harmonics
constitute an orthonormal basis like the spherical harmonics, $Y_{lm}({\bf \Omega})$, -
in the two-particle case.

After a separation of the hyperangular part in the total wave function one may obtain
equations which are equivalent to the Schr\"{o}dinger equation of a motion of 
single particle in external field.

In this approach, the centrifugal barrier of a three-body system is proportional to
the factor (K+3/2)$\cdot$(K+5/2) corresponding to the $\ell$($\ell$+1) factor
for a two-particle barrier. Usually it  is much larger than
 the two-particle barrier and is not equal to zero even for K=0. 
The derived hyperradial wavefunctions display formally  the same asymptotic behaviour
as in the 2-body case \cite{delves60}.
\vspace{-5 mm}

\subsection*{Expansion of a  decay amplitude into a series of K-harmonics}

The hyperspherical harmonics method is applied for  direct three-particle decays 
by B.V.~Danilin {\it et al.}, \cite{dan87}, 
 where the decay amplitude is suggested to be expanded in a series in the K-harmonics
basis, in analogy with the  partial wave analysis in a 2-particle case. 
The general formulae for this approach can be found in \cite{kor90}.

I will highlight the HH method for a description of energy and angular
correlations of decay fragments. 
At low energies,
 the  direct decays  should be 
determined by few components
 in the amplitude expansion which corresponds to a minimal value of hypermomentum
because of a three-particle centrifugal barrier which grows with an increase of K.
Thus, in the first approximation one may describe direct decays 
considering only few fit components with the lowest value of hypermomentum.
Then an analysis of data is simple. For example, the energy spectra of fragments
can be fitted by a superposition of few components with definite values of
quantum numbers and with a specific energy dependence each.
A weight of each component gives an information about the  norm of the respective 
configuration in the initial nucleus\footnote
{The long-range Coulomb interaction may influence the fragment distributions and
this can be taken into account in a conventional way as a final state interaction.}.

 The additional notations are given below. The decay to the three particles {\em i, j} 
 and {\em k} with 
decay energy $Q$ may be characterized by 5 kinematical variables
$ {\bf \Omega}_{i}= ({\bf \Omega}_{j-k},{\bf \Omega}_{i-jk},x_{i})$
where the quantity $x_{i}=E_{i}{\em
MQ}^{-1} (m_{j}+m_{k})^{-1}$=$E_{i}/E^{max}_{i}$, $E_{i}$ is the energy of
particle {\em i} in the c.m.\ frame of the decaying nucleus, and
$m_{i,j,k}$ are the fragment masses, ${\em M}$=$m_{i}$+$m_{j}$+$m_{k}$.
 The energy $E_{i}$
is related to the energies of the relative motion of particles {\em j}
and {\em k} ($E_{j-k}$, or $E_{x}$), and of their centre of mass and particle {\em
i} ($E_{i-jk}$, or $E_{y}$) as: $E_{i}= E_{y}(m_{j}+m_{k}){\em M}^{-1}$, and 
$E_{x}+E_{y}={\em Q}$.
The state with spin J  decays into three particles
with spins $s_{i}$, $s_{j}$ and
$s_{k}$.  To describe the final state  the following
quantum numbers are used: The {\em hypermomentum} $K=l_{x}+l_{y}+2n$
(n=0,1,2,\ldots); the orbital angular momenta ${\bf l}_{x}$ and ${\bf l}_{y}$
conjugated to the Jacobi momenta ${\bf p}_{x}$ and ${\bf p}_{y}$
and satisfying conservation of parity $\pi$ of the decaying state:
$\pi=(-1)^{l_{x}+l_{y}}$; the total orbital angular momentum
$\hat{{\bf L}}=\hat{\bf l}_{x}+\hat{\bf l}_{y}$ 
satisfying the conservation law of the total angular momentum
$\hat{{\bf J}}=\hat{{\bf L}}+\hat{{\bf S}}\;$; the total spin of all
products $\hat{{\bf S}}=\hat{{\bf s}}_{i}+\hat{{\bf s}}_{j}+\hat{{\bf
s}}_{k}$.  The total spin of any pair of
particles is ${\bf S}_{i-j}={\bf s}_{i}+{\bf s}_{j}$.

The decay amplitude of a state with spin J and its projection M 
can be expanded in a series in an orthonormal
hyperspherical harmonics basis:  
\vspace{-3 mm}

\begin{equation} \label{eq0}
F_M = \sum^{}_{KLl_{x}l_{y}}{B^{l_{x}l_{y}}_{KLS}
\cdot Y^{l_{x}l_{y}}_{KLM}({\bf p}_{x},{\bf p}_{y}) \cdot C(S_{j-k},T_{j-k})}
\end{equation}
The expansion coefficients
B$^{l_{x}l_{y}}_{KLS}$ corresponding to the hyperspherical functions
${ Y}^{l_{x}l_{y}}_{KLM} $ are
determined by the decay dynamics and give, when  squared, the
probabilities for the observed decay modes, classified according to the
sets of values of $l_{x},l_{y},K$, $L,S,S_{j-k}{\;}$. The spin-isospin weight factors
C($S_{j-k},T_{j-k}$) can be found in \cite{kor90}. Expression
(\ref{eq0}) corresponds to the simple case where the spin projections of particles are
not measured\footnote
{Under some conditions the decay amplitude may depend on the possible orientation of 
the spin of the initial state which has decayed, see \cite{boch92}.
}.

In the case of two identical particles, e.g. protons, 
the most convenient arrangement of the
Jacobi coordinates is (p--p,pp--"core"), while the set (p--"core",p--p"core")
 is close to coordinates in the c.m. system of a decaying nucleus which is
 normally used in shell-model or mean-field calculations.
Conversion in the representation of (\ref{eq0}) from one set of Jacobi coordinates
(${j-k},\!{i-jk}$) to another set (${k-j},\!{i-kj}$) is accompanied by a change of
the coefficients B$^{l_{j-k}l_{i-jk}}_{KLS}$ in accordance with the formula:
\vspace{-3 mm}

\begin{eqnarray} \label{raynal}
B^{l_{k-j}l_{i-kj}}_{KLS} = \sum_{l_{j-k},l_{i-jk}}^{}
{<l_{j-k}l_{i-jk}{\mid}l_{i-k}l_{j-ik}> 
\cdot{B^{l_{j-k}l_{i-jk}}_{KLS}}},
\end{eqnarray}
where $<...{\mid}...>$ are Raynal-Revai coefficients \cite{smor73}.

\paragraph*{Analyses of the $^6$Be, $^6$Li$^*$, $^6$He$^*$  decays into 
$\alpha$+N+N.}
The approach is tested in analyses of direct decays of A=6 nuclei. {\em Eg,}
the energy spectrum of $\alpha$-particles from $^6$Be can be
fitted by the sum of two components only, with S$_{p-p}$=0 and S$_{p-p}$=1, and
with K$_{min}$=2.
 The studied decays are governed by a single K value
in the amplitude 
expansion. Thus, the hypermomentum is confirmed to be a good
quantum number.

This interpretation is not unique in  describing  spectra of single 
$\alpha$-particles. For example, the $\alpha$-spectrum   measured
in \cite{gee77} was first fitted using the model of sequential 
emission of protons via the intermediate nucleus $^5$Li. However, the two
mechanisms predict quite different behaviour of $\alpha$--proton correlations
in a kinematical complete experiment. In particular, the three-body approach
predicts the different angular dependences of p-p correlations  with total
spins S$_{p-p}$=0 and S$_{p-p}$=1 while in the sequential model these two modes
are indistinguishable.  In the decisive kinematical complete experiment 
\cite{boch89}
where the $^6$Be decay was measured by detecting both $\alpha$-particles and
protons, the different angular distributions of the S$_{p-p}$=0 and S$_{p-p}$=1 
modes were observed directly thus confirming the 
three-body decay mechanism.

\paragraph*{Nuclear structure reflected in direct decays.}
  The fragment spectra from studied direct decays reflect the three-body 
nuclear structure. For example, the  $\alpha$+N+N correlations in A=6 nuclei
calculated in 
\cite{kuk86,dan88,dan93}
  (the fragment correlations both in space
 and in  involved angular momenta) 
 agree quantitatively with the conclusions of data analysis. 
In particular, the strong  momentum and space correlations 
between two valence protons are found in $^6$Be. These are the 
'di-proton' and 'cigar' configurations. In the first case,  two valent protons 
forms a relatively compact cluster, in the second they  are mainly  
on opposite sides relative to 
the  $\alpha$-particle. 
  The observed correlations  are  induced 
by the specific three-body quantum effect:  kinematic focusing of 
particles over momenta and in  space. This phenomenon generalizes  the 
angular 
dependance of a two-particle scattering  with $\ell\neq$0 to the 
  three-body case with K$\neq$0   and it has a universal nature. 
\vspace{-4 mm}

\section*{Direct two-proton emission from nuclei - candidates of
two-proton radioactivity}
\vspace{-2 mm}

This effect is expected to be essential in 
other nuclei with  three-body structure, {\it eg} 
the two-proton emitters $^{19}$Mg, $^{34}$Ca and $^{48}$Ni. Their estimated 
2p-decay Q-values  are less than 1.5 MeV. Since the FWHM of the  
p-p scattering spectrum
 is  $\sim$3 MeV,  the well-known mechanism of a sequential decay
via emission of $^2$He is not plausible here.
Predictions of  2p-decay modes of these nuclei done in the HH approach
are presented below. The $^{19}$Mg decay is first  described in details.

\paragraph*{Two-proton emission from $^{19}$Mg.}
According to the ref.\cite{beam99}, the main uncertainty in predictions of
$^{19}$Mg properties is given by poorly known ground-state masses of $^{19}$Mg
and $^{18}$Na: 0.9(3) MeV and 1.5(7) MeV above the $^{17}$Ne+p+p threshold, 
respectively.  
Due to the large errors in  mass estimates, two opposite
situations are possible:
i) the $^{18}$Na+p threshold is below the $^{19}$Mg ground state  
which decays by a sequential emission of protons via $^{18}$Na;
ii) the $^{18}$Na+p threshold is well above $^{19}$Mg and a
 direct two-proton emission into the $^{17}$Ne+2p channel dominates.
For the last case, I would like to
consider a direct two-proton emission in the
three-body approach which was first 
applied for the $^6$Be  data in the ref.\cite{dan87}.

To describe the decay of the $^{19}$Mg ground state which has an unknown
spin-parity, one should assume some structure of $^{19}$Mg.
%
Since $^{19}$Mg has 12 protons, two "valent" protons are likely in the
$d_{5/2}$ shell and should have the total spin S$_{p-p}$=0.  Thus, the spin-parity of
$^{19}$Mg are probably  1/2$^-$, the same as 
the $^{17}$Ne ground state.
If one assumes that the $^{19}$Mg$_{g.s.}$  has
mainly the $^{17}$Ne+p+p cluster structure,  it decays  directly 
 into the same fragments.
As the mentioned transition ends in the
$^{17}$Ne(1/2$^-$)+2p channel, the lowest value of the hypermomentum K, allowed by
momentum and parity conservation, is zero.
One may estimate the  three-particle decay width 
of the suggested $^{19}$Mg
state using the three-body model \cite{grig98} where the
"generalized R-matrix" approach is suggested. 
The ordinary R-matrix formula for decaying
states is there replaced by a similar expression $\Gamma_K$(E)=2P$_{K+3/2}$(E)$\gamma^2_K$,
where the penetrability for a three-particle decay is practically the same
function as in the R-matrix approach:  
P$_{K+3/2}$(E)=$\frac{\kappa\rho}{F^2_{K+3/2}+G^2_{K+3/2}}$.
The physical meaning of the partial reduced width $\gamma^2_K$ is the same as in the
two-body case, characterizing the spectroscopic factor for the three-particle
exit channel with a hyperradius $\rho$.  Functions $F_{K+3/2}$ and $G_{K+3/2}$ 
are calculated as
regular and irregular Coulomb functions. 
The
Wigner limit  of a reduced width is assumed.
The value of the calculation parameter, the radius of channel, 
is chosen of 16 f using an 
extrapolation of the systematic behaviour of a channel radius extracted from the fitted 
 known widths of nuclei with dominated direct three-particle decay channels:  
$^{6}$Be ($\sim$8 f), $^{10}$He ($\sim$11 f),  $^{16}$Ne
($\sim$16 f) {\em etc}, according to \cite{grig99}.

\begin{table}[thb]

\caption{Calculated widths of the $^{19}$Mg ground state for the decay energy
Q=0.89 MeV. The main component in the $^{19}$Mg wave function is assumed with the
hypermomentum K=0.  }

\begin{tabular}{c|cc}

Radius of channel, f &        Penetrability  &         Width (MeV)\\
\tableline

 14.  &                   5.40$\cdot10^{-7}$ &   3.13$\cdot10^{-6}$  \\    
 16.  &                   2.10$\cdot10^{-6}$ &  0.139$\cdot10^{-4}$  \\  
 18.  &                   7.27$\cdot10^{-6}$  & 0.542$\cdot10^{-4}$   \\  
 20.  &                  0.226$\cdot10^{-4}$ &  0.000187 \\     
\end{tabular}

\caption{The same as in Table 1, 
except Q-values.}
\begin{tabular}{c|cc|cc}

Channel     &  Penetrability  & Width (MeV) &  Penetrability  &  Width (MeV)\\ 
radius, f & Q=0.54 MeV & Q=0.54 MeV & Q=1.24 MeV & Q=1.24 MeV\\
\tableline
 14. &   3.84$\cdot10^{-11}$  & 1.73$\cdot10^{-10}$ &0.595  &    0.000499 \\    
 16. &  1.65$\cdot10^{-10}$  & 8.57$\cdot10^{-10}$ &0.673  &    0.00198 \\     
 18. &  6.40$\cdot10^{-10}$  & 3.72$\cdot10^{-9}$ & 0.731  &    0.00679 \\    
 20. & 2.24$\cdot10^{-9}$   & 1.45$\cdot10^{-8}$ & 0.773  &    0.0203\\     
\end{tabular}

\caption{The same as in Table 1, except dominating hypermomentum value, K. }
\begin{tabular}{c|cc|cc}
Channel     &  Penetrability  & Width (MeV) &  Penetrability  &  Width (MeV)\\ 
radius, f & K=2 & K=2 & K=4 & K=4\\
\tableline

 14. & 2.35$\cdot10^{-8}$   &  1.36$\cdot10^{-7}$ & 1.95$\cdot10^{-10}$ &  1.13$\cdot10^{-9}$\\     
 16.  & 1.18$\cdot10^{-7}$   & 7.87$\cdot10^{-7}$ & 1.42$\cdot10^{-9 }$ &  9.45$\cdot10^{-9}$\\     
 18. &5.13$\cdot10^{-7}$   & 3.82$\cdot10^{-6}$ & 8.45$\cdot10^{-9 }$ &   6.30$\cdot10^{-8}$\\     
 20. & 1.94$\cdot10^{-6}$   & 0.161$\cdot10^{-4}$ & 4.23$\cdot10^{-8 }$ &  3.50$\cdot10^{-7}$\\    
\end{tabular}

\end{table}

In Tables 1--3  the results of $^{19}$Mg width calculations 
are shown for varied values of
the ground state position respective to the $^{17}$Ne+p+p threshold, the channel
radius $\rho$ and the hypermomentum K, \cite{grig99a}.
As one may see, the estimated width of the $^{19}$Mg ground
state is of 10 eV  
judging purely by assumption that the K$_{min}$=$\ell_{p-p}+\ell_{Ne-pp}$=0 
configuration
dominates  the $^{19}$Mg wave function. However, due to  Pauli principle, such a
component should be suppressed because the valent
protons have to be  in the $d_{5/2}$ shell 
(or $\ell_{p-Ne}=\ell_{p-pNe}$=2).  The configuration with next hypermomentum, K=2, 
(the estimated width
$\sim$1 eV) does not match to the assumed $d_{5/2}$ shell structure as
well.  Finally, the configuration with K=4  is
allowed by the Pauli principle  
(because the assumed ($d_{5/2}$)$^2$ wave in $^{19}$Mg overlaps by 90\%
with the K=4, $\ell_{p-p}$=$\ell_{Ne-pp}$=0 component, using eq.(2))
and the respective width of
the ground state is of 0.01 eV.  The last value corresponds to a very long
life-time, of 10$^{-14}$ s, and thus the decay could be classified as 
a radioactivity phenomenon.

\begin{figure}[th!]
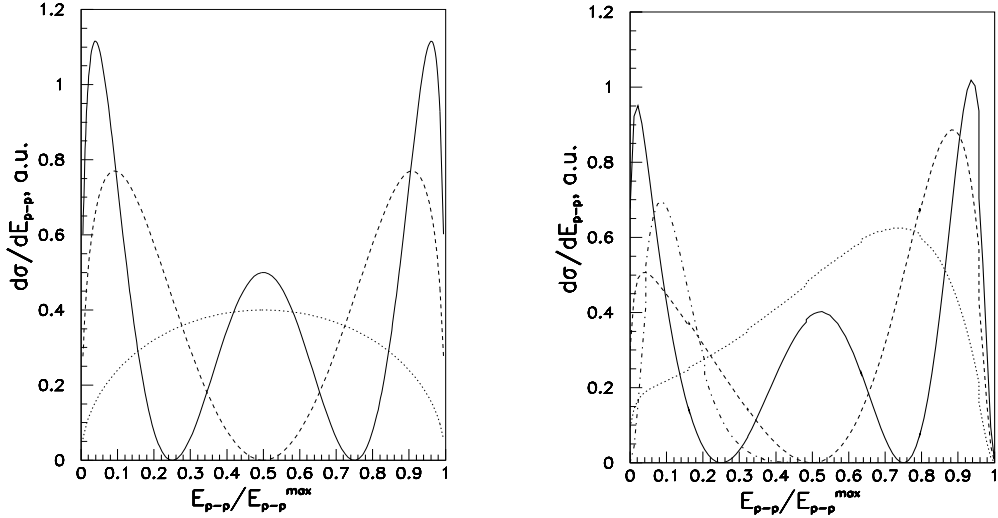


\hspace{-0.3 cm}
{\epsfig{file=mg19,  width=7.cm, height=8cm}}
\vspace{-8.0cm}

\hspace{7.cm}
{\epsfig{file=mg19fsi,  width=7.cm, height=8cm}}

\caption{On left:  possible E$_{p-p}$ spectra from the direct decay
$^{19}$Mg(1/2$^-$)$\rightarrow^{17}$Ne+p+p where
$\ell_{p-p}$=$\ell_{Ne-pp}$=0.  The p-p energy is
given as a fraction of the maximum E$_{p-p}$ value.
The solid, dashed and dotted curves correspond to
the decay modes with K=4, 2 and 0, respectively.   On right:  the same
components as shown to the left, and in addition the Coulomb repulsion of
the fragments is taken into account. The dot-dashed curve
is the result of diproton model.}

\label{fig1}
\end{figure}

On the basis of the assumed structure of $^{19}$Mg one may predict properties of its
direct three-particle decay (see details in Appendix).
If the K=0 and K=2 components in
$^{19}$Mg are suppressed by the Pauli principle completely, the decay amplitude and 
spectra of the fragments should only be defined
by the K$_{min}$=4 configuration with $\ell_{p-p}=\ell_{Ne-pp}$=0.  
For example, the corresponding E$_{p-p}$ spectrum\footnote
{The $^{17}$Ne spectrum can easy be
obtained from the E$_{p-p}$ spectrum using the formulae
E$_{Ne}$=$\frac{2}{19}$E$_{Ne-pp}$ and E$_{Ne-pp}$+E$_{p-p}$=Q.  } 
is shown in
fig.\ref{fig1}, on left, by a solid curve.  Dashed and dotted curves are the results
of a calculation of the suppressed decay modes with K=2 and 0, respectively.

The long-range Coulomb repulsion must be taken into account when a more
realistic approximation is required.  In this case the final state interaction model
can be used. The full decay amplitude $F$ is then factorized as
$|F|^2=|F_3|^2{\cdot}{\mid{F_{FSI}}\mid}^2$, where $|F_3|$ is the 
three-particle decay amplitude, and
$|F_{FSI}|$ - the  final state interaction factor, see Appendix.
 The calculated E$_{p-p}$ spectra from $^{19}$Mg
 are shown in fig.\ref{fig1}, on right: the dotted, dashed, solid curves 
 corresponds to
the K=0,2,4 components, respectively.  As one can see, the final state
interaction influences significantly the E$_{p-p}$ spectrum of the K=0 mode
only.
The considered configurations in $^{19}$Mg  have very different
 probabilities to be observed in its decay.  One may quantitatively compare the
 calculated penetrabilities of the K=0,2,4 modes for the $^{19}$Mg direct
 three-particle decay,  which values are 
 2$\cdot10^{-6}$, $10^{-7}$, $10^{-9}$, respectively (see Tables 1--3,
 the row with the radius of
 channel of 16 f).

Thus, there could be three exotic modes of the 2p-decay of $^{19}$Mg.  
First, if the K=0
component is not forbidden strictly by the Pauli principle and its admixture to
the dominating K=4 configuration is more than 0.1\%, the
K=0 mode should mainly be observed,
 with weak p-p correlations as shown in
fig.\ref{fig1} by dotted curves. Second, if the admixture of the suppressed K=2
component in $^{19}$Mg  is more than 1\%, the
strong p-p correlations (like in the di-proton model) should be 
 detected  as well as strong p-p
anticorrelations, see dashed curves in fig.\ref{fig1}.  Third, if the Pauli
principle suppresses the mentioned components strongly,  very
exotic p-p correlations with three peaks in E$_{p-p}$ spectrum should 
appear, see
solid curves in fig.\ref{fig1}. Combinations of these three basic decay modes are
possible as well.

One should also consider the decay branch 
$^{19}$Mg(1/2$^-$)$\rightarrow^{17}$Ne(1/2$^-$)+p+p, 
 with L=1. In this decay with K$_{min}$=2,in order
to conserve the momentum and parity, two protons have to be with
S$_{p-p}$=1 and the relative orbital momenta - with
$\ell_{p-p}$=$\ell_{Ne-pp}$=1. This component corresponds 
with 100\% probability to the $\ell_{p-Ne}$=$\ell_{p-pNe}$=1 configuration, 
which agrees with
two valent protons  being in the p-shell of $^{19}$Mg and
with S$_{p-p}$=1. This contradicts to the
assumed ($d_{5/2}$)$^2$  structure of $^{19}$Mg and therefore
 such a component is unlikely  compete with the  decay modes 
considered above.

\vspace{-0.8 cm}
\begin{figure}[thb]
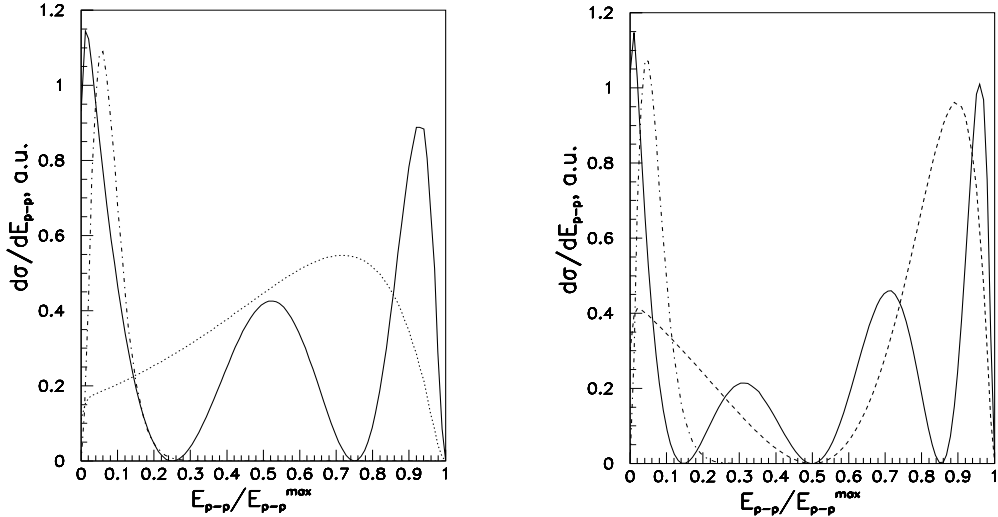


\hspace{-0.3 cm}
{\epsfig{file=ca34fsi,  width=7.cm, height=8cm}}
\vspace{-8.0cm}

\hspace{7.cm}
{\epsfig{file=ni48fsi,  width=7.cm, height=8cm}}

\caption{  The expected E$_{p-p}$ spectra from the direct decays:
$^{34}$Ca(0$^+$)$\rightarrow^{32}$Ar+p+p with
$\ell_{p-p}$=$\ell_{Ar-pp}$=0, on left; $^{48}$Ni(0$^+$)$\rightarrow^{46}$Fe+p+p with
$\ell_{p-p}$=$\ell_{Fe-pp}$=0, on right. The p-p energy is
given as a fraction of the maximum E$_{p-p}$ value. 
The curves are the same as in fig.1, except the solid curve, to the right,
which corresponds to the decay mode with K=6. 
 The Coulomb repulsion of
fragments is taken into account in all spectra. }

\label{fig2}
\end{figure}

\paragraph*{Decay modes and life-time estimates of $^{34}$Ca and $^{48}$Ni.}
The decay properties of $^{34}$Ca are expected to be similar to those of
$^{19}$Mg. The $^{34}$Ca ground state  is calculated to be bound respective to the
single proton emission (Q$_p$=--0.9 MeV) and unbound respective to
the $^{32}$Ar+p+p decay with Q$_{2p}$=0.75 MeV \cite{brown91}. According to 
a conventional shell-model, $^{34}$Ca
has a complete proton $d$-shell. Therefore its spin-parity is 0$^+$ and one
may apply all considerations valid for the $^{19}$Mg case. The only significant
differences are the higher Coulomb barrier and the smaller Q$_{2p}$-value which result
in much larger estimated life-times: $\sim10^{-10}$ s, $10^{-9}$ s, 
$5 \cdot 10^{-7}$ s for the dominating K=0,2,4 components, 
respectively \cite{grig99a}. 
The corresponding 2p-decay modes are shown as E$_{p-p}$ spectra 
in fig.2, on left.

The recently discovered (see J.~Giovinazzo et al., Proc. of PROCON99) 
double-magic nucleus $^{48}$Ni(0$^+$) with a complete proton
$f_{7/2}$-shell is estimated to be
bound respective to the 1p-emission (Q$_p$=--0.46 MeV) and it may decay into the
$^{46}$Fe+p+p channel with Q$_{2p}$=1.36 MeV \cite{brown91}. Since two valent
protons are expected to be in the $f$-shell,  the lowest hypermomentum
allowed by the Pauli principle is K=6. The corresponding single term in the HH expansion of 
the decay
amplitude is Y$^{00}_{600}$
(see notation in Appendix, eq.(4))
and the respective impressive 2p-decay correlations 
are shown by solid curve in fig.2, on right. If a small admixture  of the 
$(p_{1/2})^2$-wave, of 1\%,
is present in $^{48}$Ni, then the K=2 component should dominate in its decay (like
in the $^{19}$Mg   case) with
the E$_{p-p}$ spectrum shown by the dashed curve in fig.2, on right.
The result of diproton model is shown by dash-dotted curve for a comparison.
The  high Coulomb barrier for a 2p-decay causes 
a very large life-times:  $\sim 10^{-6}$ s and $\sim 10^{-4}$ s for the 
K=2 and K=6 components, respectively \cite{grig99a}. 
\vspace{-2 mm}

\section*{Summary}
\vspace{-1 mm}

Two-proton decay is  a three-body problem in a case of non-sequential, or
direct two-proton emission.
As direct decays may be studied by expanding a decay amplitude 
in a series of hyperspherical harmonics functions,
the fragment spectra  can be fitted by a few components determined by 
relative  orbital momenta of fragments and
 a single, minimal value of hypermomentum.

Strong correlations of fragments, observed in  direct decays of the A=6 nuclei,
reflect exotic three-body configurations in  the mother
nuclei.
These modes are induced by a three-body quantum effect of general nature: 
kinematic focusing of
fragments over momenta and in space due to K$\neq$0.

Because of this three-body phenomenon, strong p-p correlations
 are expected in direct two-proton emission of other nuclei, {\em eg}
two-proton radioactivity candidates $^{19}$Mg, $^{34} $Ca, $^{48} $Ni.
The life-time and decay properties of these nuclei 
 considered in a three-body approach  depend
strongly on the mass and the structure of  ground states. 
Exotic decay modes, {\em eg} two-proton radioactivity with
strongly oscillating p-p correlations, may appear making these drip-line
nuclei attractive objects for experimental studies.
\vspace{-4 mm} 

\section*{Acknowledgements}
\vspace{-2 mm} 

The author wishes to gratefully acknowledge the collaboration of
colleagues.
In particular,
L.~Grigorenko has made several contributions to this work.
The support of the Gesellschaft f\"{u}r Schwerionenforshung mbH, 
the German Federal Minister for Education and Research
(BMBF) under Contract 06 DA 820 and the PROCON99 Organizing Committee 
is acknowledged.
\vspace{0.5 cm}

{\large \bf Appendix. Direct 2p-decays of 0$^+$ nuclear states.}
\vspace{0.5 cm}

Explicit formulae for a simultaneous emission
of two protons from the ground state of $^{19}$Mg are presented here, 
namely, for the
transitions with $\Delta J^{\pi}$ of 0$^+$ and 1$^+$.

The probability of the $^{19}$Mg decay  into the $^{17}$Ne+2p channel
 can be derived in analogy with the
$^6$Be 2p-decay, \cite{boch89}.  The $^{17}$Ne 
spectrum  differs from the $^6$Be  case by
normalization coefficients only and can be fitted by the expression
\begin{eqnarray} \label{eq250}
 \frac{\partial^2 P}{\partial E_{Ne}\partial \Omega_{ Ne}} =
 \frac{19}{16{\pi}Q} \sqrt{x_{Ne}(1-x_{Ne})} \cdot \mid F\mid^2 \cdot \mid F_{FSI}\mid^2,
\end{eqnarray}
where $x_{}$=E$_{Ne}$/E$_{Ne}^{max}$.
For the direct two proton decay with
$\Delta{J}^{\pi}$=$0^+$, the amplitude approximation ${F}$ has few components, e.g.
the expansion with the  lowest possible {\em hypermomentum} values of 0, 2, 4
and 6 is:
\begin{eqnarray} \label{eq251}
 & F & (p\!-\!p,Ne\!-\!pp)\!  \sim \!
     B^{00}_{000}Y^{00}_{000}  + B^{00}_{200}Y^{00}_{200} + 
    B^{00}_{400}Y^{00}_{400} + B^{00}_{600}Y^{00}_{600}  \nonumber \\
  & = &  B^{00}_{000}   + B^{00}_{200}(2x_{}-1) + 
  B^{00}_{400}(16x_{}^2-16x_{}+3) + B^{00}_{600}(2x-1)(8x(x-1)+1).
\end{eqnarray}

The decay amplitude (\ref{eq251}) expressed via  other Jacobi
coordinates ($p$-$Ne$,$p$-$pNe$) is: 
\begin{eqnarray} \label{eq77}
F & = &
B^{00}_{000}[Y^{00}_{000}] + B^{00}_{200}[0.99Y^{11}_{200}+0.05Y^{00}_{200}]   
  +  B^{00}_{400}[0.94Y^{22}_{400}
  -0.09Y^{11}_{400}-0.33Y^{00}_{400}] \nonumber \\
  & + & B^{00}_{600}[0.89Y^{33}_{600}
  -0.04Y^{22}_{600}-0.45Y^{11}_{600}+0.02Y^{00}_{600}]
\end{eqnarray}
where new norms are calculated  using the Raynal-Revai transformation 
(\ref{raynal}) of the expansion coefficients
$B^{00}_{K00}$ from (\ref{eq251}).

As the coordinates ($p$-$Ne$,$p$-$pNe$) are almost the same  as 
the proton coordinates in the $^{19}$Mg c.m. system, the equation (\ref{eq77})
can be used for estimates of the single-proton configurations in $^{19}$Mg
corresponding to the respective p-p modes in the eq.(\ref{eq251}).
In particular, the first component in the eq.(\ref{eq251}), 
with K=0, $\ell_{p-p}$=$\ell_{Ne-pp}$=0, with the 100\% probability corresponds
to the $\ell_{p-Ne}$=$\ell_{p-pNe}$=0 component in eq.(\ref{eq77}) 
which matches  the valent protons being
in the $s$-shell of $^{19}$Mg. The second term in (\ref{eq251}), when K=2, 
with the (0.9985)$^2$=0.997  probability coincides
with the $\ell_{p-Ne}$=$\ell_{p-pNe}$=1 component in eq.(\ref{eq77})
which represents valent protons
in the $p$-shell. And the last term  in (\ref{eq251}), with K=4, 
has the (0.94)$^2$=0.884  probability to
overlap with the $\ell_{p-Ne}$=$\ell_{p-pNe}$=2 component which 
matches  the valent protons in the $d$-shell.


In eq.~(\ref{eq250}), the final state interaction factor F$_{FSI}$ was
used (as in the Migdal-Watson model applied in the $^6$Be decay
case) to take into account the Coulomb repulsion of charged particles
and the p-p attraction in a S=$\ell_{pp}$=0 wave:
$ \!
 {\mid{F_{FSI}}\mid}^2 = P_{Ne-p_1} 
  P_{Ne-p_2}{\cdot}\mid\Phi_{p-p}\mid^2
 \! $
where $P_{i-j}(E_{i-j})={(F^2_0+G^2_0)}^{-1}$ is the Coulomb
penetration factor for particles {\em i,j} and $\mid\Phi_{p-p}\mid^2$
is a p-p interaction factor taken in the effective range expansion
\cite{phil64}.
 

\vspace{-2 mm}




\begin{thebibliography}{99}
\vspace{-1 mm}


\bibitem{boch89} O.V.~Bochkarev et. al., Nucl.~Phys. {\bf A505}, 215-222 (1989). 
\vspace{-0.05 cm}
\bibitem{gee77}D.F.~Geesaman et al., Phys. Rev. {\bf C15}, 1835-1852 (1977).
\bibitem{boch92}O.V.~Bochkarev et al., Sov. J. Nucl. Phys. {\bf 55}, 955-969 (1992).
\bibitem{delves60} L.M.~Delves, Nucl.~Phys. {\bf 20}, 275-308 (1960). 
\vspace{-0.05 cm}
\bibitem{dan87} B.V.~Danilin et. al.,  Sov. J. Nucl. Phys. {\bf 46}, 225 (1987).  
\vspace{-0.05 cm}
\bibitem{kor90}A.A.Korsheninnikov, Sov. J. Nucl. Phys. {\bf 52}, 1304-1315 (1990).
\bibitem{kuk86} V.I.~Kukulin et.al., Nucl. Phys., {\bf A453}, 365-382  (1986). 
\vspace{-0.05 cm}
\bibitem{dan88}B.V.~Danilin et.al., 
Sov.~J.~Nucl.~Phys. {\bf 48} (1988) 766;  {\bf 49} (1988) 217;
 {\bf 53} 71-87 (1991)  
\bibitem{dan93}B.V.~Danilin and M.V.~Zhukov, Phys. At. Nucl. {\bf 56}, 460-475 (1993).
\bibitem{beam99}D.~Beamel et al., "Search for two-proton emission from $^{19}$Mg",
proposal to the GANIL program committee, 1999.
\bibitem{grig98}L.V.~Grigorenko, "Electromagnetic and weak interactions
in light exotic nuclei", Ph.D. thesis, 1997, Chalmers University of Technology,
Geteborg, ISBN 91-7197-553-5.
\bibitem{grig99}L.V.~Grigorenko, I.G.~Mukha and M.V.~Zhukov,
"Analysis of a $\beta$-delayed multi-particle emission in a few-body approach:
the $^9$Li test",
in preparation.
\bibitem{grig99a}L.V.~Grigorenko, private communication. 
\bibitem{phil64}R.J.N.~Phillips, Nucl.Phys. {\bf A53}, 650-662 (1964).
\bibitem{smor73}Ya.A.~Smorodinskij and V.D.~Efros, 
Sov.~J.~Nucl.~Phys. {\bf 17},  107-123 (1973).
\bibitem{brown91}B.~Alex Brown, Phys. Rev. {\bf C43}, R1513-R1517 (1991).

\end{thebibliography}
\end{document}